\documentclass[a4paper]{jpconf}
\usepackage{graphicx}
\begin{document}
\title{Metal-insulator transition in the hybridized two-orbital Hubbard model revisited}

\author{Y. N\'u\~nez-Fern\'andez and K. Hallberg}
\address{Centro At{\'o}mico Bariloche and Instituto Balseiro, CNEA and CONICET, 8400 Bariloche, Argentina}
\ead{yurielnf@gmail.com}

\begin{abstract}
In this work we study the two-orbital Hubbard model on a square lattice in the presence of hybridization between nearest-neighbor orbitals and a crystal-field splitting. We use a highly reliable numerical technique based on the density matrix renormalization group to solve the dynamical mean field theory self-consistent impurity problem. We find that the orbital mixing always leads to a finite local density states at the Fermi energy in both orbitals when at least one band is metallic. When one band is doped, and the chemical potential lies between the Hubbard bands in the other band, the coherent quasiparticle peak in this orbital has an exponential behavior with the Hubbard interaction $U$.
\end{abstract}

\section{Introduction}

The Mott-Hubbard metal-insulator transition (MIT) remains a highly interesting problem in correlated systems due to its relevance in many families of materials interesting physical properties and with current and potential applications. \cite{Mott, imada1} The development of the dynamical mean field theory (DMFT) \cite{review} has triggered a great deal of theoretical research leading to important progress in the understanding of the electronic structure in these systems. 

At the core of the DMFT method is the calculation of Greens functions of an interacting quantum impurity for which the fermionic bath has to be determined self-consistently until convergence of the local Green function and the local self-energy is reached. Recent advances in numerical techniques for the resolution of multiorbital quantum impurity problems have made possible the implementation of the Dynamical Mean Field Theory (DMFT) in model Hamiltonians and  Density-Functional-based methods for the calculation of strongly correlated materials in a realistic way.\cite{EPLreview}  

One of the most reliable impurity solvers is the Density Matrix Renormalization Group (DMRG)\cite{white1, scholl, karen1, garciadmft, garciadmft2, yuriel} which allows for the calculation of the density of states directly on the real axis and at zero temperature (or with a very small imaginary offset). This methods complements Monte Carlo based impurity solvers such as the CTQMC \cite{ctqmc1,CTQMC} which suffer from the fermionic sign problem (particularly for hybridized models), from approximate analytic continuation of the Green's functions to the real energy axis and are useful for finite temperatures. It is comparable, though, to the NRG-based impurity solver.\cite{bulla, vollhardt, Zitko}

In this paper we use the DMFT with the DMRG as the impurity solver to study the hybridized two-orbital Hubbard model and the effect of the inter-orbital hopping and the crystal-field splitting in the Mott transition. 

\section{Model and method} 

We consider the following model:

\begin{equation}
H=\sum_{ij\alpha\beta}t_{ij}^{\alpha\beta}c_{i\alpha}^{\dagger}c_{j\beta}-\mu\sum_{i}n_{i}\mbox{,}+U\sum_{\alpha}n_{i\alpha\uparrow}n_{i\alpha\downarrow}
\label{eq:H}
\end{equation}
where $c^{\dagger}$ creates an electron, $i,j$ are sites indices on the square lattice, $\alpha,\beta=1,2$ are the orbital indices, spin indices are implicit, and we consider the paramagnetic solution. Here we define the on-site and hopping operators in the orbital basis:

\begin{equation}
{\bf t}_{0}=\left(\begin{array}{cc}
\Delta & 0\\
0 & -\Delta
\end{array}\right),\mbox{ } {\bf t}_{x}=\left(\begin{array}{cc}
t & t'\\
t' & t
\end{array}\right),\; {\bf t}_{y}=\left(\begin{array}{cc}
t & -t'\\
-t' & t
\end{array}\right),
\end{equation}
where the subindices indicate the on-site, $x$ and $y$ hopping matrices, $\Delta $ is the crystal field splitting and $t'$ is the inter-orbital hopping parameter.
We will consider $t=0.25$ as the unit of energy.

Defining the matrix $T(k)$ in the reciprocal space $T(k)={\bf t}_{0}+2{\bf t}_{x}\cos k_{x}+2{\bf t}_{y}\cos k_{y}-\mu {\bf I}$ and the self-energy $\Sigma(k,z)$, $z=\omega+i\eta$ we can write the Green's function as:
\begin{equation}
G(k,z)=\left[z\,I-T(k)-\Sigma(k,z)\right]^{-1}
\end{equation}

It can be shown that, in the half-filled case, the Hamiltonian (\ref{eq:H}) is invariant under the transformation $c_{i\alpha}^{\dagger}\rightarrow(-1)^{x_{i}+y_{i}+\alpha}c_{i\bar{\alpha}}$ and in the square lattice the Green's function is diagonal, $G_{12}(\omega)=G_{21}(\omega)=0$:

\begin{equation}
G(z)=\left(\begin{array}{cc}
G_{11}(z) & 0\\
0 & G_{22}(z)
\end{array}\right),\label{eq:Gmatrix}
\end{equation}
where  $G_{22}(z)=G_{11}(-z)$ for half filling. 

We use the DMFT to obtain the electronic densities of states in this model. This technique approximates the momentum dependent self-energy by a local quantity
$\Sigma(k,z)\approx\Sigma(z)$. The hybridization used in this method has the same structure as (\ref{eq:Gmatrix}) and this means that both orbitals are decoupled, having independent hybridization baths, which simplifies the calculation. We define the effective impurity Hamiltonian $H_{imp}$ taking part in the DMFT equations as:

\begin{equation}
H_{imp}=\sum_{\alpha}t_{0}^{\alpha \alpha}c_{0\alpha}^{\dagger}c_{0\alpha}+U\sum_{\alpha}n_{0\alpha\uparrow}n_{0\alpha\downarrow}+H_{b}\mbox{,}
\end{equation}
where the non-interacting bath term $H_{b}$ is:

\begin{equation}
H_{b}=\sum_{\alpha i}\lambda_{i}^{\alpha}b_{\alpha i}^{\dagger}b_{\alpha i}+\sum_{\alpha i}v_{i}^{\alpha}\left[b_{\alpha i}^{\dagger}c_{0\alpha}+H.c.\right]\mbox{,}
\end{equation}
where $b_{\alpha i}^{\dagger}$ represents the creation operator for the site $i$ associated to the bath site $\alpha$.

The DMFT iterations are the following:

\begin{description}
\item [{(i)}] Define $\Sigma(z)=0$,
\item [{(ii)}] Calculate the Green's function at the impurity site for each orbital and the hybridization: 
\begin{equation}
G_{\alpha\alpha}(z)=\frac{1}{N}\sum_{k}G_{\alpha\alpha}(k,z)=\frac{1}{N}\sum_{k}\left[z\,I-T(k)-\Sigma(z)\right]_{\alpha\alpha}^{-1}
\end{equation}

\begin{equation}
\Gamma_{\alpha\alpha}(z)=z-t_{0}^{\alpha\alpha}-\Sigma_{\alpha\alpha}(z)-G_{\alpha\alpha}^{-1}(z)\mbox{.}
\end{equation}

\item [{(iii)}] Find an approximation $\tilde{\Gamma}_{\alpha\alpha}(z)$  for the hybridization $\Gamma_{\alpha\alpha}(z)$ via the parameters
$v_{i}^{\alpha}$ and $\lambda_{i}^{\alpha}$:
\begin{equation}
\tilde{\Gamma}_{\alpha\alpha}(z)=\sum_{i}\frac{\left|v_{i}^{\alpha}\right|^{2}}{\omega-\lambda_{i}^{\alpha}}\mbox{.}
\end{equation}

\item [{(iv)}] Calculate the Green's function $G_{\alpha\alpha}(z)$ at site ``$0$'' of $H_{imp}$ using DMRG:
\item [{(v)}] Calculate the self-energy: 
\begin{equation}
\Sigma_{\alpha\alpha}(z)=z-t_{0}^{\alpha\alpha}-G_{\alpha\alpha}^{-1}(z)-\tilde{\Gamma}_{\alpha\alpha}(z)\mbox{.}
\end{equation}
Return to \textbf{(ii)} until convergence.
\end{description}

 \section{Results}

\begin{figure}[h]
\begin{center}
\includegraphics[width=15cm]{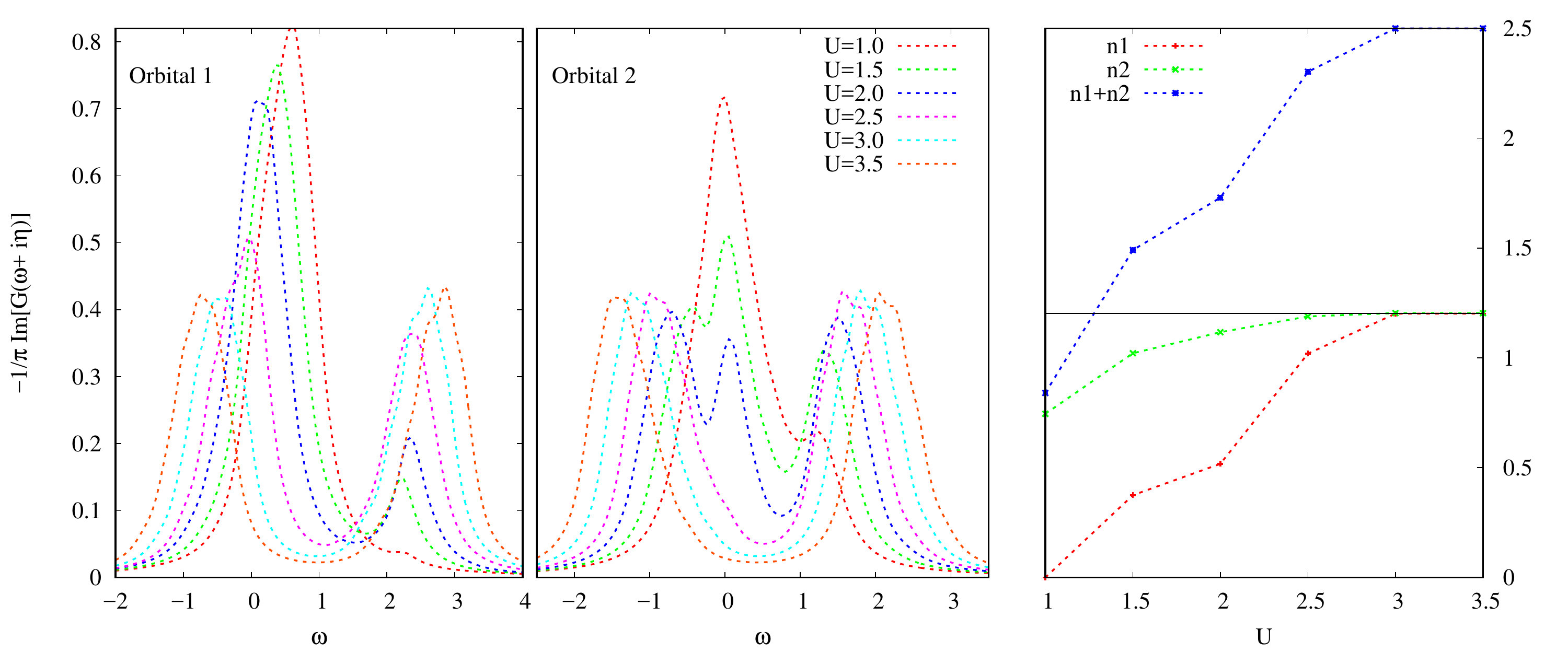}
\end{center}
\caption{\label{figure1} DOS for the hybridized two-band Hubbard model with a finite crystal-field splitting $\Delta=0.3$, $t'=0.15$ and $\mu=-0.7$, increasing $U$ showing a simultaneous metal-insulator transition in orbitals 1 and 2.  Right: Orbital fillings vs. $U$.}
\end{figure}

Using the method described above, we obtained the densities of state (DOS) of model (\ref{eq:H}) for several representative parameters.
In Fig. \ref{figure1} we show the behavior of both orbitals for different values of the interaction $U$ and a finite crystal field splitting $\Delta$. Also shown is the particle filling of each orbital vs. $U$. We find that, due to the interorbital hopping $t'$, the lower-lying orbital (orbital 2) develops a large weight at the Fermi energy. Both orbitals transition to an insulating state upon increasing $U$ as seen also in the filling values, approaching the half-filled situation ($n=1$) simultaneously.

\begin{figure}[h]
\begin{center}
\includegraphics[width=15cm]{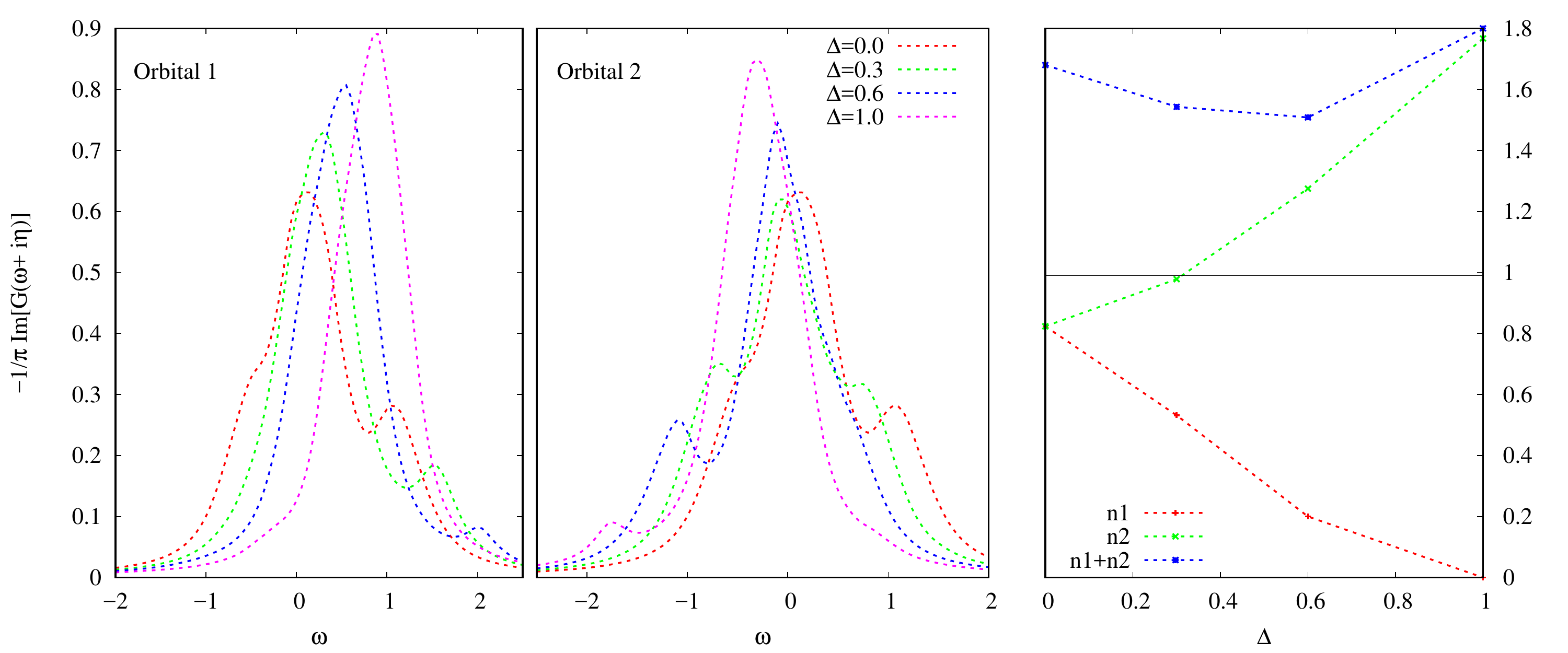}
\end{center}
\caption{\label{figure2} DOS for $U=1$, $t'=0.15$ and $\mu=-0.3$,  changing $\Delta $ showing a transition from a metal to a band insulator in orbitals 1 and 2. Right: The corresponding orbital fillings.}
\end{figure}

For small interactions we find a non-correlated metal-band insulator transition by varying the crystal-field splitting, as shown in Fig. \ref{figure2}.

\begin{figure}[h]
\begin{center}
\includegraphics[width=15cm]{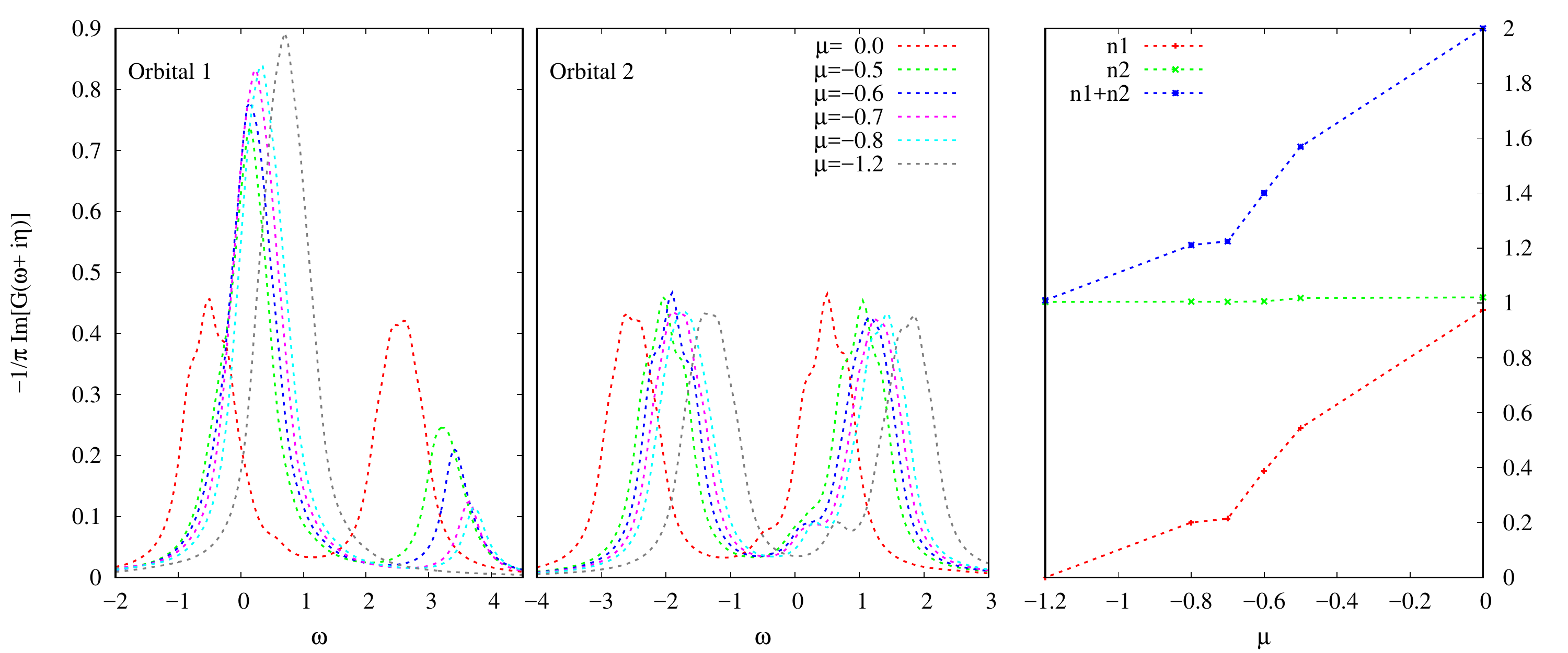}
\end{center}
\caption{\label{figure3} Same as Fig. 2 but for $U=3$,  $\Delta=1$, for different chemical potentials and their corresponding fillings (right).}
\end{figure}

In Fig. \ref{figure3} we vary the chemical potential and see that, for these parameters, orbital 2 remains half-filled while band 1 empties by increasing the absolute value of $\mu$. A finite DOS at the Fermi energy in orbital 1 always leads to a finite DOS in orbital 2, seen here as a small peak at zero energy.

\begin{figure}[h]
\begin{center}
\includegraphics[width=15cm]{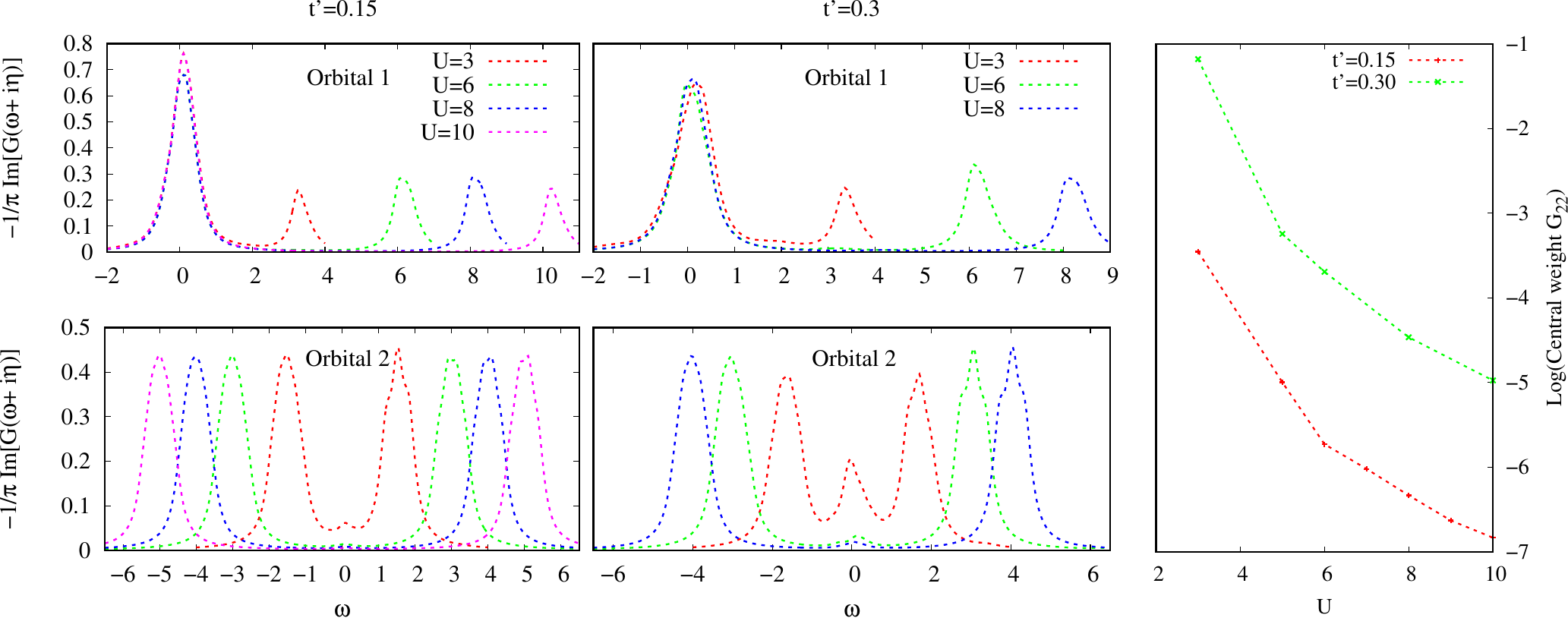}
\end{center}
\caption{\label{figure4} DOS for fixed $\Delta=0.75$ and $\mu=-0.75$ for which the Fermi energy lies in the lower Hubbard band for orbital 1 and between both Hubbard bands in orbital 2 for two values of inter-orbital hybridization $t'$. Note the shift in energy values between the left and middle panels. Right: Quasiparticle weight in orbital 2 vs. $U$.}
\end{figure}

It is interesting to analyze the effect on orbital 2 of a large density of states in orbital 1. We find that, when the Fermi energy lies in the lower Hubbard band of the upper orbital (orbital 1), it causes a finite DOS in the lower orbital at the Fermi energy, even for large values of $U$, see Fig. \ref{figure4}. As a consequence, orbital 2 does not have a Mott metal-insulator transition and there is no selective MIT.\cite{kotliarnoMIT} 

We measured the integral of the peak at the Fermi energy of orbital 2 (which is proprtional to the quasiparticle weight $Z_2$). We find that, as long as the Fermi energy lies within the Hubbard band in orbital 1, the quasiparticle weight in orbital 2 will be finite.  All other parameters fixed, $Z_2$ decreases exponentially with $U$ like in the Kondo problem (see  Fig. \ref{figure3}), not showing a metal-insulator transition for any value of $U$. The larger the inter-orbital hopping, the larger the DOS at the Fermi energy in orbital 2 (see Fig. 4).


\section{Conclusions}
By using a reliable zero temperature technique which is free from fermionic signs and delivers the Green functions on the real energy axis directly to solve the DMFT impurity problem, we calculate the density of states of the hibridized two-band Hubbard model on a square lattice. We find that for finite inter-orbital mixing, there is a finite density of states at the Fermi energy of both bands even for large interactions, as long as one band remains metallic. The metal-insulator transition, if it occurs, happens simultaneously for both orbitals.  By doping one of the bands and fixing the chemical potential in between the Hubbard bands in the other correlated orbital we find that the density of states in this second orbital remains finite even for large interactions. Here we do not find any MIT and the correlated metallic quasiparticle peak decreases exponentially with with the Hubbard interaction $U$.

\subsection{Acknowledgements}
We acknowledge support from projects PICT 2012-1069 and PICT 2016-0402 from the Argentine ANPCyT and PIP 2015-2017 11220150100538CO (CONICET).  This work used the Extreme Science and Engineering Discovery Environment (XSEDE), which is supported by National Science Foundation grant number ACI-1548562.

\end{document}